\documentclass[english,reprint,amsmath,amssymb,aps,prl]{revtex4-2}
\usepackage[T1]{fontenc}
\usepackage[latin9]{inputenc}
\usepackage{color}
\usepackage{amsmath}
\usepackage{graphicx}
\usepackage{braket}

\makeatletter
\usepackage{babel}

\usepackage{xcolor}
\usepackage{hyperref}

\hypersetup{
    colorlinks=true,
    linkcolor={red!50!black},
    citecolor={green!50!black},
    urlcolor={blue!50!black}
}

\makeatother

\usepackage{babel}
\begin{document}
\title{Bogolyubov excitons as a microscopic origin of two-level systems}
\author{Andrey Grankin, Victor Galitski}
\affiliation{Joint Quantum Institute, Department of Physics, University of Maryland,
College Park, MD 20742, USA}
\begin{abstract}
Two-level systems (TLS) are a leading source of decoherence and
dielectric loss in Josephson-junction qubits, yet their microscopic
origin remains unresolved. We propose an intrinsic, purely electronic
TLS candidate: subgap bound states of Bogolyubov quasiparticles. We show that,
in a model of a conventional superconductor with electron-electron attraction
in the $s$-wave channel and repulsion in higher-angular-momentum channels,
the latter produce an effective attraction between Bogolyubov
quasiparticles, forming bound states. These Bogolyubov excitons resemble Bardasis--Schrieffer excitons but are driven by repulsive interactions. Bulk Bogolyubov excitons are not easily detectable through single-particle tunneling or other conventional probes, but we show
that surface excitons couple to an external electric field and behave as
two-level systems. We examine an idealized model of a bulk superconductor
proximity-coupled to a two-dimensional repulsive metal, which could represent
an external metallic or semiconductor layer or an underscreened region of the
superconductor. The two levels correspond to the absence and presence of a
single exciton, which carries an electric dipole moment and exhibits an
avoided crossing with a resonator, as observed for TLS. Because this mechanism
requires no defects and cannot be eliminated by screening or annealing, it
suggests that TLS-like excitations may be intrinsic to superconductors.\end{abstract}
\maketitle
The main experimental signatures conventionally attributed to TLS
in superconducting heterostructures are: (\textbf{i})~TLS are quantum systems,
with  lifetimes of order 100$ns$~\citep{Cooper2004,Lisenfeld2010,Shalibo2010}.
(\textbf{ii}) They couple strongly to electric fields and behave as microscopic
electric dipoles \citep{Martinis2005,Lisenfeld2019}. (\textbf{iii})~Individual
TLS appear as microwave resonances producing avoided level crossings
in qubit spectroscopy \citep{Simmonds2004,Palomaki2010,Cooper2004}.
(\textbf{iv})~Their dielectric response is saturable, with characteristic temperature-
and microwave-drive dependence \citep{OConnell2008,Kumar2008Temperature,Martinis2005}.
(\textbf{v})~Finally, the TLS appear in both the Josephson tunnel barriers
and the superconductor surfaces and interfaces \citep{Lisenfeld2019,Bilmes2022},
with experimental results being consistent with TLS located
in a nm-thick surface layer \citep{Gao2008Surface}.  Below we refer to these as signatures (\textbf{i})-(\textbf{v}). 
\begin{figure}
\includegraphics[scale=0.3]{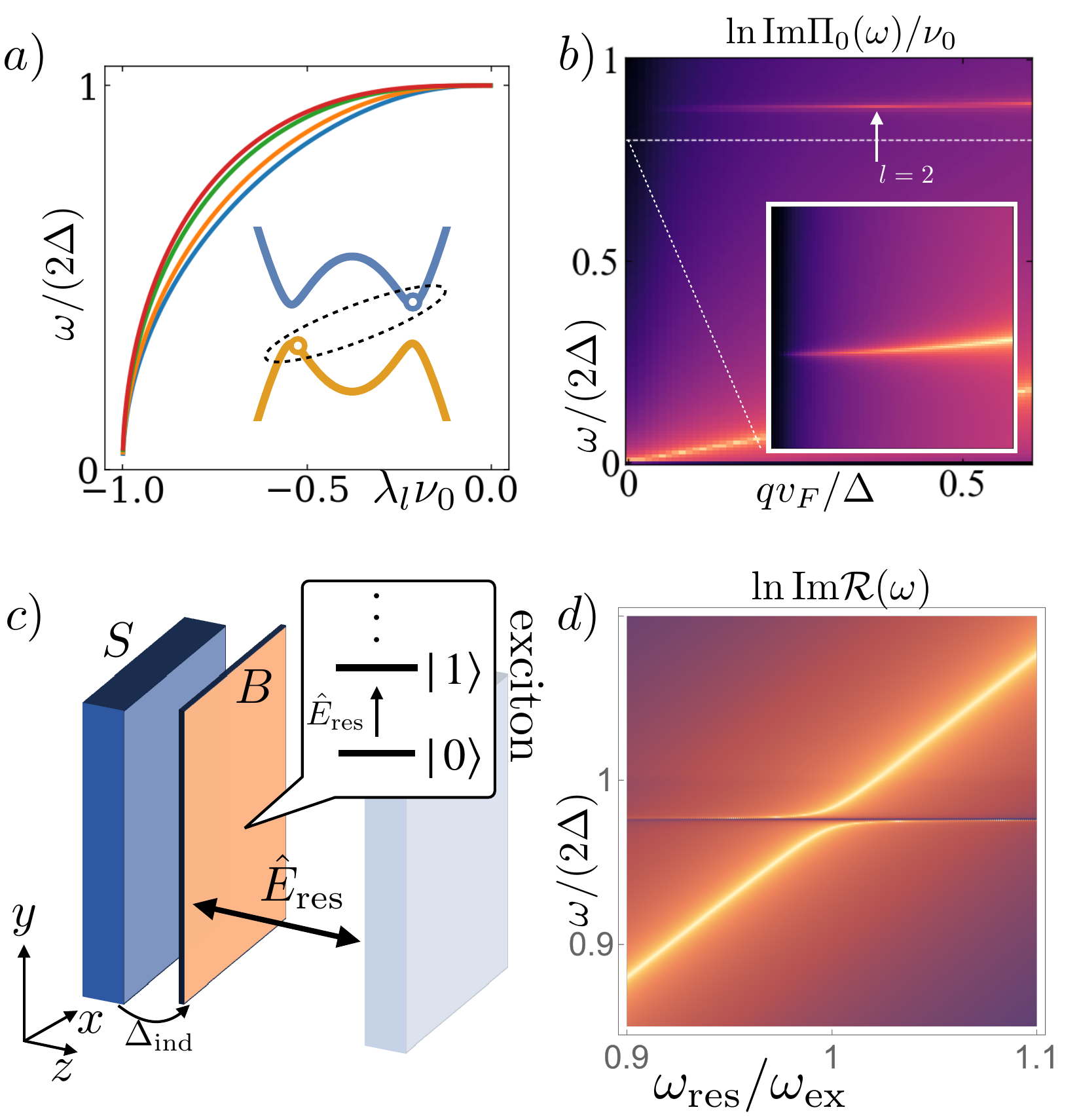}

\caption{Repulsion-induced Bogolyubov exciton formation in a two-dimensional superconductor.
(a) Exciton energy as a function of the interaction strength in $l$-th channel for $\lambda_0\nu_0=0.1$ (blue), $0.2$ (orange), $0.4$ (green), $0.5$ (red). Bogolyubov quasiparticle bands, with quasielectrons and quasiholes interacting via the Coulomb interaction and forming an excitonic collective degree of freedom, are shown in the inset. 
(b) Low-${\bf q}$ density-density response $\text{Im}\Pi_{\bf{q}}(\omega)/\nu_0$ of a gate-screened superconductor for  $\lambda_2\nu_0=-0.5$. High-energy excitonic modes emerge at finite $\bf q$. The inset shows a zoom on frequencies above $1.6\Delta$. (c,d): Schematic illustration of the boundary inhomogeneity resonator setup. (c) A bulk parent superconductor
(S) is tunnel-coupled to the boundary layer (B). S acts as a source of Cooper pairs
characterized by $\Delta_{\text{ind}}$. The second superconductor forming
a resonator with S is shown as a pale blue rectangle and is capacitively coupled,
thereby inducing a voltage between S and B. Exciton-resonator coupling produces an avoided crossing. (d) Spectral function of the resonator ${\cal R}(\omega)$
as a function of resonator frequency $\omega_{\text{res}}$. Avoided
crossing is observed when $\omega\approx\omega_{\text{ex}}$. Parameters
for this simulation are $\gamma_{\text{res}}=10^{-3}\omega_{\text{res}},$
$\lambda_{0}\nu_{0}=-0.25$, $\Delta_{\text{ind}}=E_{0}/10$. The
dimensionless coupling strength is chosen to match the reasonable
avoided crossing scale $(dE^{\text{ZPF}})^{2}\nu_{0}{\cal A}/\omega_{\text{res}}\approx5\times10^{-6}$. Since the most visible resonances are Dirac-$\delta$-function-like, we omit the colorbars.}

\label{Fig1}
\end{figure}

The microscopic origin of TLS in superconducting circuits has not
been established conclusively \citep{Mueller2019}. The standard tunneling
model \citep{Phillips1972} describes them phenomenologically as atoms
or small groups of atoms tunneling between nearly degenerate local
configurations in an amorphous dielectric. Candidate microscopic
realizations include hydrogen-related defects and surface OH rotors
in alumina \citep{Holder2013}, delocalized oxygen atoms in AlOx barriers
\citep{DuBois2013}, non-stoichiometric oxide defects such as oxygen
vacancies \citep{PhysRevApplied.22.024035}, and, for some coherent
defects, trapped quasiparticles in shallow subgap states \citep{deGraaf2020}.
%In this Letter, we propose an alternative scenario where subgap resonances
%emerge from many-body nature of superconducting gap fluctuations at
%the boundary of a macroscopic superconducting region experiencing
%strong Coulomb repulsion.

In this Letter, we propose intrinsic many-body excitonic resonances in superconductors as a possible microscopic origin of two-level systems in superconducting devices. In contrast to excitons induced by a subdominant attractive pairing channel \cite{gassner2024light,allocca2019cavity,bardasis1961excitons,vaks1962collective}, the modes considered here originate from purely repulsive Coulomb interactions, which bind Bogolyubov quasielectrons and quasiholes as shown in Fig.~\ref{Fig1}~(a). We first demonstrate how such modes arise in a minimal model of a two-dimensional intrinsic superconductor and show that higher-angular-momentum excitons appear in the finite-momentum density-density response. We then extend our analysis and demonstrate the emergence of localized excitons in a boundary inhomogeneity model which accounts for the underscreened Coulomb interactions at the interface. By incorporating this model into a superconducting resonator geometry as shown in Fig.~\ref{Fig1}(c) we then propose a smoking gun experiment allowing for a direct probe of Bogolyubov excitons in experiment. In this case, the Bogolyubov excitons are predicted to become  dipole-active either at finite in-plane momentum or, at ${\bf q}=0$ in the presence of weak rotational-symmetry breaking. This setup therefore exhibits several experimental signatures TLS defined above: (\textbf{i})~sharp resonances below the quasiparticle continuum, (\textbf{ii}) a finite dipole moment in the surface-inhomogeneity model, and (\textbf{iii}) coupling to a superconducting resonator, producing characteristic avoided crossings in the spectral response. Furthermore, surface underscreening is expected to enhance exciton binding near the boundary, providing a natural localization mechanism consistent with (\textbf{v}).

To demonstrate the emergence of repulsion-induced electron-hole bound states, we consider a simplified model of a two-dimensional superconductor where the electrons interact through a potential $V_{\bf{q}}$. 
The Hamiltonian is:

\begin{align}
H & =\sum_{{\bf k},\sigma}\xi_{k}\hat{\psi}_{{\bf k},\sigma}^{\dagger}\hat{\psi}_{{\bf k},\sigma} - \sum_{{\bf q}}V_{{\bf q}}\hat{\rho}_{{\bf q}}\hat{\rho}_{-{\bf q}}\label{eq:H0}
\end{align}
where the electronic density operator is defined as $\hat{\rho}_{{\bf q}}={\cal A}^{-1/2}\sum_{{\bf k},\sigma}\hat{\psi}_{{\bf k+q},\sigma}^{\dagger}\hat{\psi}_{{\bf k},\sigma}$
with $\hat{\psi}_{{\bf k},\sigma}(\hat{\psi}_{{\bf k},\sigma}^{\dagger})$
being the fermionic annihilation operators of momentum ${\bf k}$,
spin $\sigma$, ${\cal A}$ is the area of the system, and $V_{{\bf q}}$
is the interaction strength that accounts for both phonon-induced electron
attraction and Coulomb repulsion. The electronic dispersion is defined
as $\xi_{k}=k^{2}/2m-\mu$ where $m$ is the electronic mass and $\mu$
is the chemical potential. In the following, we expand the interaction
potential in angular harmonics as follows 
\begin{equation}
\label{Vl}
V_{{\bf k-k'}}=\sum_{l}\lambda_{l}\, e^{il \left(\phi_{{\bf k}}-\phi_{{\bf k'}}\right)}
\end{equation}
We consider the leading $s$-wave channel to be attractive, i.e. $\lambda_{0}>0$. Some or all other channels can be repulsive. The key finding is that they still lead to an effective binding of Bogolyubov quasiparticles. 

We now apply mean-field approximation for the Hamiltonian Eq.~\eqref{eq:H0}.
We define the conventional Nambu spinors $\Psi=\{\psi_{{\bf k},\uparrow},\psi_{-{\bf k},\downarrow}^{\dagger}\}^{\text{T}}$
with the corresponding imaginary-time Green's function being
\begin{equation}
\hat{{\cal G}}_{{\bf k}}^{-1}(i\epsilon_{n})=i\epsilon_{n}-\left(\begin{array}{cc}
\xi_{k} & \Delta\\
\Delta & -\xi_{k}
\end{array}\right),\label{eq:G}
\end{equation}
where $\epsilon_{n}=(2n+1)\pi T,n\in\mathbb{Z}$, $T$ is the temperature
and $\Delta$ is the interaction induced anomalous self energy defining
the superconducting gap. Assuming perfectly symmetric circular Fermi
surface, it is determined via the BCS self-consistency equation:

\begin{align}
1 & =\frac{T}{\cal A}\lambda_{0}\sum_{{\bf k},n}\frac{1}{\epsilon_{n}^{2}+\Delta^{2}+\xi_{k}^{2}},\label{eq:delta}
\end{align}
where we defined Nambu Pauli matrices as $\hat{\tau}_{i}$. In deriving
Eq.~\eqref{eq:delta}, we assumed the Fermi surface is perfectly
s-wave symmetric. %We solve this equation self-consistently by imposing a high-energy cut-off $E_{0}$. 

\begin{figure}
\includegraphics[scale=0.35]{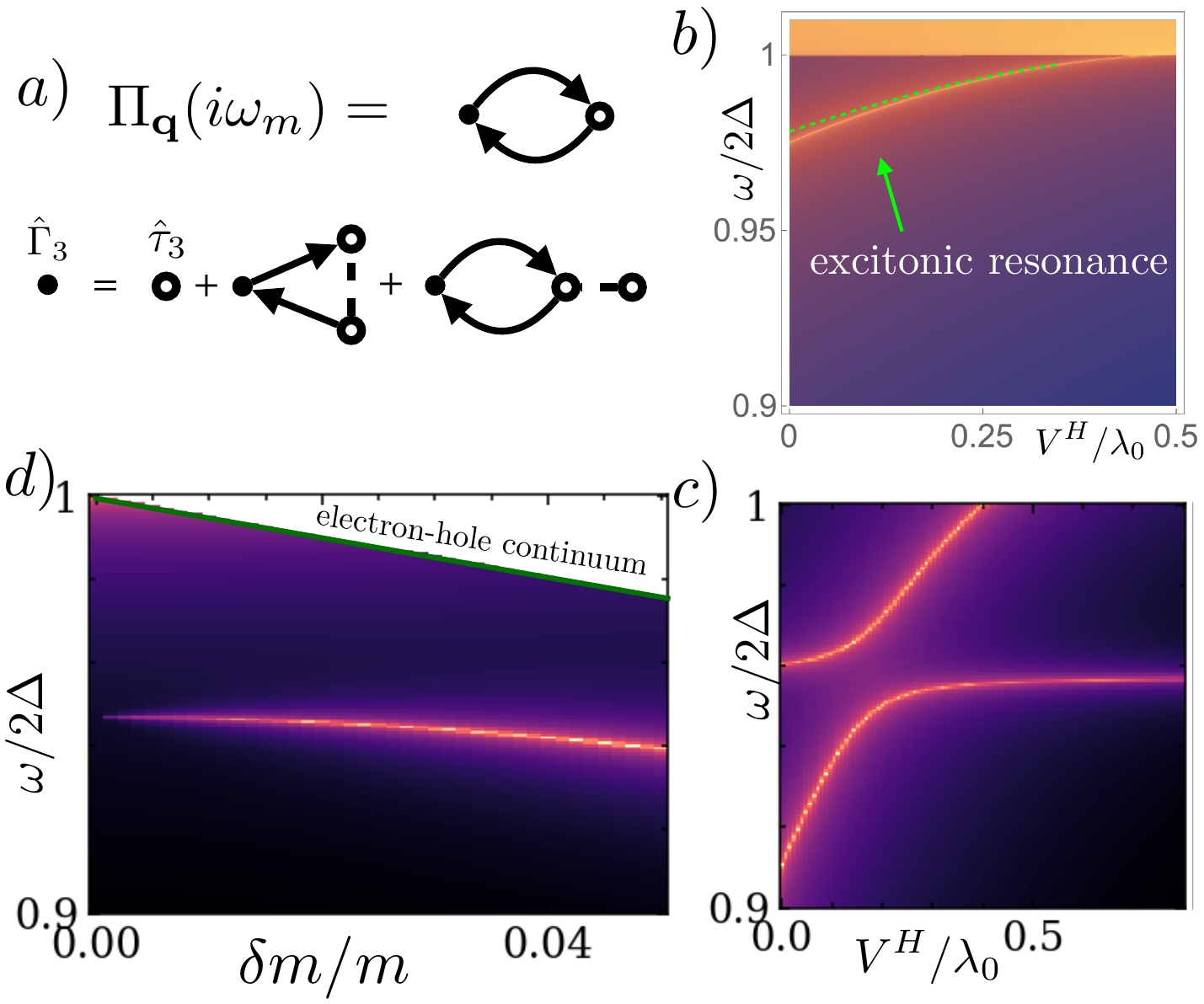}

\caption{(a) Diagrammatic
representation of vertex corrections in the charge sector. Empty circles
correspond to the bare charge vertex $\hat{\tau}_{3}$ while filled
circles denote $\hat{\Gamma}_{3}$ according to Eq.~\eqref{eq:BS}.
Solid arrows represent fully renormalized Green's functions Eq.~\ref{eq:G}
while dashed lines represent interaction terms. (b) Charge response
polarization function $\text{Im}\Pi_{{\bf{q}}=0}(\omega)/\nu_0$ of the S-B junction as a function of the
frequency $\omega/2\Delta$, where $\Delta=\Delta_{\text{ind}}+\delta$
and the Hartree charging energy $V^{H}$ for $\lambda_{0}\nu_0=-0.3$.
Exciton resonance is visible as sharp resonance below the quasiparticle
continuum $2\Delta$. Analytical estimation of exciton energy according
to Eq.~\ref{eq:omega_ex} is shown with the green dashed curve. (c) Polarization function in the presence of a weak rotational-symmetry-breaking perturbation $\delta m/m$ and $V^H\nu_0=-0.3$. 
(d) $\operatorname{Im}\Pi_0(\omega)/\nu_0$ as a function of $V^H$ for fixed $\delta m/m=0.025$. 
The $s$-wave exciton, Eq.~\eqref{eq:omega_ex}, is shifted by $V^H$ to higher energies, and an avoided crossing is observed when the energies of the $s$- and $d$-wave excitons coincide.  In (c) and (d), we assume the following parameters: $\lambda_0\nu_0=-0.5$ and $\lambda_2\nu_0=-0.4$. The white region in (d) denotes the modified particle-hole continuum in the presence of symmetry breaking. }

\label{Fig2}
\end{figure}

We now determine the collective excitation spectrum on top of the
mean-field solution Eq.~(\ref{eq:G}, \ref{eq:delta}) enforcing gauge-invariance \cite{schrieffer2018theory,maiti2017conservation}. To this end, we consider the density-density polarization function
of the electron gas $\Pi({\bf q},i\Omega_{m})=-\int e^{i\Omega_{m}\tau}\langle\hat{\rho}_{{\bf q}}(\tau)\hat{\rho}_{-{\bf q}}(0)\rangle$,
where $\Omega_{m}=2m\pi T$ with $m\in\mathbb{Z}$. Within the mean-field
theory, polarization function can be expressed through the fully renormalized
Green's functions Eq.~\eqref{eq:G} and the bare and density vertices
$\hat{\tau}_{3}$ and $\hat{\Gamma}_{3}$ as follows

\begin{equation}
\Pi_{{\bf q}}(i\Omega_{m})=\frac{T}{{\cal A}}\sum_{{\bf k},n}\text{Tr}\hat{{\cal G}}_{{\bf k+\frac{q}{2}}}(i\epsilon_{n+m})\hat{\Gamma}_{3}({\bf k})\hat{{\cal G}}_{{\bf k-\frac{q}{2}}}(i\epsilon_{n})\hat{\tau}_{3}.\label{eq:Pi}
\end{equation}
$\hat{\Gamma}_{3}$ includes the collective excitations in superfluids
that are generally required to be taken into account to ensure the
charge conservation \citep{schrieffer2018theory}. It can be shown
that $\hat{\Gamma}_{3}$ \citep{schrieffer2018theory} obeys Bethe-Salpeter
equation diagrammatically shown in Fig.~\ref{Fig2}(a). 
\begin{align}
\hat{\Gamma}_{3}({\bf k}) & =\hat{\tau}_{3}+\sum_{{\bf k}'}V_{{\bf k-k'}}{\cal K}(\hat{\Gamma}_{3}({\bf k}'))-V_{{\bf q}}^{H}\hat{\tau}_{3}\Pi_{{\bf q}}(i\Omega_{m}),\label{eq:BS}\\
{\cal K}(\hat{\Gamma}_{3}) & \equiv\frac{T}{{\cal A}}\sum_{n'}\hat{\tau}_{3}\hat{{\cal G}}_{{\bf k'+\frac{q}{2}}}\left(i\epsilon_{n'+m}\right)\hat{\Gamma}_{3}\hat{{\cal G}}_{{\bf k'-\frac{q}{2}}}(i\epsilon_{n'})\hat{\tau}_{3},
\end{align}
where we partially omitted variables of $\hat{\Gamma}_{3}({\bf k})$
for shortness. For the non-retarded interaction employed throughout
this work, both equations \eqref{eq:Pi} and \eqref{eq:BS} can be
solved analytically at ${\bf q}=0$. The analytic continuation to real
frequencies in Eq.~\eqref{eq:Pi} amounts to the simple replacement $i\Omega_{m}\rightarrow\omega+i0^{+}$.
The last term in Eq.~\eqref{eq:BS} describes an additional Hartree
vertex correction \citep{schrieffer2018theory} that is responsible
for coupling of collective modes to electromagnetic field. Importantly,
conventionally $V_{{\bf k-k'}}$ represents the interaction that is
screened by the electron gas itself while $V_{{\bf q}}^{H}$ is bare
interaction. Here for simplicity we restrict ourselves to the dominant Coulomb
interaction and assume it is additionally gate-screened $V_{{\bf q}}^{H}=-2\pi e^{2}(1-e^{-qd})/q\approx-2\pi e^{2}d$,
where $d$ is the distance to the gate\citep{grankin2023interplay}. 

At ${\bf q}=0$ the collective excitation spectrum is determined by
the poles of Eq.~\eqref{eq:BS}, which can be found analytically by performing
the angular harmonics decomposition of the Bethe Salpeter equation.
In case of rotationally-symmetric $s$-wave Fermi surface, the pole
in $l\neq0$ channel can be found analytically (see SM) at weak coupling and
is given by 
\begin{equation}
\omega_{\text{ex}}^{l\neq0}\approx2\Delta-\frac{\pi^{2}\Delta\lambda_{l}^{4}\nu_{0}^{4}\log^{2}\frac{2E_{0}}{\Delta}}{4\left[ \log\left(\frac{2E_{0}}{\Delta}\right)(\lambda_{l}\nu_{0}-1)\lambda_{l}\nu_{0}+1\right]^{2}},\label{eq:lneq}
\end{equation}
where $E_0$ is the high-energy cutoff. We identify these exciton-like resonances below the quasiparticle continuum
as repulsion-induced Bogolyubov excitons \citep{bardasis1961excitons}.
We note that the $l=0$ pole is associated with charge conservation in
this case and is always located at zero frequency as required by Ward's
identity \citep{schrieffer2018theory}. We also note that for the
same reason these higher angular momentum fluctuations do not contribute
to the response function Eq.~\eqref{eq:Pi} at ${\bf q}=0$. The $l\neq0$ resonances can still be identified by assuming
the bare vertex in Eq.~\eqref{eq:BS} has momentum dependence $\hat{\tau}_{3}\rightarrow\hat{\tau}_{3}e^{i\phi_{k}l},$
where $\phi_{k}$ is the angle of the ${\bf k}$ vector with respect
to the $x$ axis. The resulting analytical estimate of the exciton energy is shown in Fig.~\ref{Fig1}~(a)
as a function of the repulsion strength $\lambda_{l}$. 

The angular-momentum basis is also convenient for computing the response at finite center-of-mass momentum ${\bf q}$. In this case, we solve the Bethe-Salpeter equation Eq.~\eqref{eq:BS} numerically and compute the modified response function Eq.~\eqref{eq:Pi}. We truncate the angular-momentum expansion at $l=2$. The result is shown in Fig.~\ref{Fig1}(b). As expected, the response vanishes at $q=0$. In contrast, at finite $q$, higher-angular-momentum modes contribute to $\Pi_{\bf q}(\omega)$.

We now extend our formalism and consider a different scenario where the Bogolyubov excitons emerge at the boundary of a superconductor. We also show how they can be probed through capacitive coupling to a superconducting resonator. Specifically, we consider the geometry shown in Fig.~\ref{Fig1}(c), where one superconducting island of the resonator, denoted by (S), is tunnel-coupled to the two-dimensional boundary layer (B) described by Eq.~\eqref{eq:H0}.  We treat the
superconductor as a source of Cooper pairs with a constant momentum-
and frequency-independent rate $\Delta_{\text{ind}}$. Our formalism Eqs.~(\ref{eq:BS},\ref{eq:G}) is extended straightforwardly to this scenario with the only difference being that  the superconducting
gap in Eq.~\eqref{eq:G} is given by $\Delta=\Delta_{\text{ind}}+\delta$,
where $\delta$ is the intrinsic anomalous self-energy, obeying $\delta=T{\cal A}^{-1}\sum_{{\bf k},n}\lambda_{0}\Delta/(\epsilon_{n}^{2}+\Delta^{2}+\xi_{k}^{2})$. 
By performing a nearly identical analysis using Eq.~\eqref{eq:Pi},
we find a similar exciton spectrum for $l\neq0$ (See SM). For attractive interaction $\lambda_0>0$, the $s$-wave channel
becomes gapped, representing a Josephson plasmon dressed by the interactions.
At weak tunneling rate $\Delta_{\text{ind}}$, we find: $\omega_{J}^{l=0}=2\sqrt{\delta\Delta_{\text{ind}}}\sqrt{1+\frac{1-2V^{H}\nu_{0}}{\lambda_{0}\nu_{0}}}$.
As expected, the global charging energy $\left|V^{H}\right|$ increases the plasma
frequency to higher values \citep{grankin2024extended}. 

Interestingly, if our boundary layer is purely repulsive in all channels due to the weaker screening, i.e. $\lambda_{0}<0$,
and does not support superconductivity on its own, we also find a
resonance that can be interpreted as an $s$-wave exciton. Indeed,
assuming at weak coupling $\lambda_{0}<0$ we find: 

\begin{equation}
\omega_{\text{ex}}^{l=0}\approx2\Delta-\frac{\pi^{2}\Delta(\delta\lambda_{0}\nu_{0}+2V^{H}\nu_{0}\Delta_{\text{ind}})^{2}}{4\left[\delta\lambda_{0}\nu_{0}+(1+2V^{H}\nu_{0})\Delta_{\text{ind}}\right]^{2}}\label{eq:omega_ex}
\end{equation}
Furthermore, the sharply-defined $s$-wave resonance exists provided
that the condition $\left|\lambda_{0}\right|/|V^{H}|>2\left|\Delta_{\text{ind}}/\delta\right|$
is satisfied. This results from the fact that the $s$-wave charge
fluctuations are a hybrid of plasmon and exciton and can be pushed to
higher energies. The competition between different terms ($\lambda_{0}\delta$
and $V^{H}\Delta_{\text{ind}}$) depends on many factors, including
screening properties on both $S$ and $B$. We also note that regardless
of the nature of superconductivity in the boundary, the finite-angular
momentum exciton responses are given by the expression similar to Eq.~\eqref{eq:lneq}
and depend mostly on the gap value and coupling strength in the corresponding
channel. Since the Coulomb
interaction is enhanced near the boundary, these excitons can be more
strongly bound, providing a mechanism for their surface localization.

As in the isolated-superconductor scenario discussed in Fig.~\ref{Fig1}(b), coupling to higher-angular-momentum excitons in the proximitized setup can be mediated by electric-field components with finite in-plane momentum $q$, which are generally present in realistic devices. In the boundary case, however, these excitons can also be probed through the voltage response across the S-B junction even at ${\bf q}=0$, provided rotational symmetry is weakly broken. To illustrate this mechanism, we consider a simple model with a weak anisotropy in the electron mass,
$m_x=m+\delta m$ and $m_y=m-\delta m$, with $\delta m/m\ll1$. As shown in Fig.~\ref{Fig2}(c,d), higher-angular-momentum excitons appear in the voltage response and persist even for large Hartree shifts $V^H$. 

%\begin{figure}
%\includegraphics[scale=0.25]{Fig3}
%
%\caption{Schematic illustration of the resonator setup. (a) Bulk parent superconductor
%(S) tunnel-coupled to boundary layer (B) acting as a source of Cooper pairs
%characterized by $\Delta_{\text{ind}}$. The second superconductor forming
%a resonator with $S$ is shown as a pale blue rectangle and is coupled capacitively,
%inducing voltage between $S$ and $B$.  Avoided crossing in exciton-resonator
%coupling. (b) Spectral function of the resonator ${\cal R}(\omega)$
%as a function of resonator frequency $\omega_{\text{res}}$. Avoided
%crossing is observed when $\omega\approx\omega_{\text{ex}}$. Parameters
%for this simulation are $\gamma_{\text{res}}=10^{-3}\omega_{\text{res}},$
%$\lambda_{0}\nu_{0}=-0.25$, $\Delta_{\text{ind}}=E_{0}/10$. The
%dimensionless coupling strength is chosen to match the reasonable
%avoided crossing scale $(dE^{\text{ZPF}})^{2}\nu_{0}{\cal A}/\omega_{\text{res}}\approx5\times10^{-6}$.}
%
%\label{Fig3}
%\end{figure}

We now consider capacitive coupling between our system and
the resonator mode as shown in Fig.~\ref{Fig1}(c). Focusing on
the simplest single-mode scenario, the ${\bf q}=0$ case, and quantizing
the resonator field, we define the dipolar coupling to our excitonic
layer (Eq.~(\ref{eq:H0})) as 

\begin{equation}
H_{\text{dip}}=dE^{\text{ZPF}}(\hat{a}+\hat{a}^{\dagger})\sum_{{\bf k}}\Psi_{{\bf k}}^{\dagger}\hat{\tau}_{3}\Psi_{{\bf k}},\label{eq:Hdip}
\end{equation}
where $E^{\text{ZPF}}$ is the zero-point electric field fluctuation
amplitude of the resonator mode $\hat{a}$ having frequency $\omega_{\text{res}}$
with the corresponding Hamiltonian $H_{\text{res}}=\omega_{\text{res}}\hat{a}^{\dagger}\hat{a}$.
$V_{\text{res}}$ denotes the mode volume, and $\epsilon$ is the dielectric
susceptibility. Integrating-out the electronic gas in Eqs.~\eqref{eq:H0},
\eqref{eq:Hdip} and using the full renormalized polarization function
Eq.~\eqref{eq:Pi} we get the response of the combined resonator-electron
system. 

\begin{equation}
{\cal R}(\omega)=\frac{-2\omega_{\text{res}}}{2\omega_{\text{res}}(dE^{\text{ZPF}})^{2}{\cal A}\Pi_{{\bf q}=0}(\omega)+(\omega_{\text{res}}^{2}-(\omega+i\gamma_{\text{res}})^{2})}\label{eq:R}
\end{equation}
where we have included the phenomenological lifetime of the resonator $\gamma_{\text{res}}$.
Near the resonance $\omega\approx\omega_{\text{ex}}$ polarization
function behaves as $\Pi_{{\bf q}=0}(\omega)\approx\nu_{0}\frac{Z_{\text{ex}}}{\omega+i\eta-\omega_{\text{ex}}}$,
where $Z_{\text{ex}}$ is the exciton residue that depends on microscopic
parameters ($\Delta_{\text{ind}},\delta$, $\lambda_{0}$, $\ldots$). We note that such behavior is identical to the response of TLS to the oscillating electric field in linear response regime, as discussed in SM. In the exciton language, the TLS is formed by the vacuum
$|0\rangle$ and the single-exciton Fock state $|1\rangle$, as illustrated in
Fig.~\ref{Fig1}(c).
Combined resonances for $\omega_{\text{res}}\approx\omega_{\text{ex}}$
are given by $\omega_{\pm}=\omega_{\text{res}}\left(1\pm\sqrt{\alpha}\right)$,
where $\alpha=(dE^{\text{ZPF}})^{2}\nu_{0}{\cal A}Z_{\text{ex}}/(\omega^2_{\text{res}})$.
Evaluating Eq.~\eqref{eq:R} numerically, we observe an avoided crossing
as a function of resonator frequency, as shown in Fig.~\ref{Fig1}(d).
The avoided crossing persists for a wide range of coupling strengths
and the resonator decoherence rate $\gamma_{\text{res}}$. 

We now discuss the experimental energy scales relevant to our proposal.
Typical microwave TLS observed in superconducting circuits have transition
frequencies in the range $4$-$12\,{\rm GHz}$ \citep{Lisenfeld2019,Simmonds2004,Palomaki2010},
 below the bulk aluminum quasiparticle threshold, $2\Delta_{{\rm Al}}\simeq2\pi\times90{\rm GHz}$.
If the relevant gap near the boundary is comparable to the bulk value, Fig.~\ref{Fig1}(a)
would imply that a strong repulsive interaction in the higher-angular-
momentum channels is required to push the exciton into the microwave
range. We emphasize, however, that the static contact interaction used
above is only a minimal parametrization of the screened Coulomb interaction.
The full screened interaction is frequency dependent, and its Fermi-surface
projection can be strongly enhanced, or even singular, at finite frequencies
relevant for exciton binding \citep{grankin2023interplay}.  A more
microscopic treatment of this interaction is left for future work. 

Our boundary inhomogeneity model provides a natural way to resolve this scale
mismatch. 
In the weak-repulsion regime the Bogolyubov exciton lies below the
corresponding local quasiparticle threshold, $\omega_{{\rm ex}}\lesssim2\Delta$,
so a strongly reduced minigap in a disordered, granular, or weakly
proximitized region can place the resonance directly in the microwave
window. Moreover, spatial inhomogeneity of the minigap can localize these excitonic modes,
producing multiple discrete TLS-like resonances rather than a single
extended collective mode. Such localization may also enhance the nonlinearity
of the response: in a finite localization volume, Pauli phase-space
filling and exciton-exciton interactions can enhance the effective
anharmonicity of the excitonic mode and therefore increase its saturability
compared with the extended collective-mode limit \citep{keldysh2024collective,huang2024nonbosonic}.

In conclusion, we propose a many-body collective electronic origin of TLS in superconducting heterostructures based on Bogolyubov excitons. We find that repulsive interactions in higher-angular-momentum channels can bind Bogolyubov quasiparticles into sharp subgap collective modes. When these modes acquire a charge component, they develop an electric dipole response and produce avoided crossings with electromagnetic resonators. In particular, we show that crystalline anisotropy, and finite-momentum electric fields can activate otherwise dark angular-momentum channels, yielding a family of sharp subgap resonances. In contrast to defect-based proposals \cite{faoro2006quantum}, our scenario can also be implemented artificially, for example by forming a junction between a bulk superconductor and a metal with controlled screening. This provides a tunable platform for exploring many-body electronic excitations as a possible source of TLS-like behavior in superconducting devices.

\begin{acknowledgments}
This work was primarily supported by the U.S. Department of Energy, Office of Science Basic Energy Sciences under
Award No. DE-SC0001911 (fundamental physics of superconductors). The authors  also acknowledge support by DARPA HR00112490310 (connection to superconducting quantum devices). 
\end{acknowledgments}

\appendix
%dummy comment inserted by tex2lyx to ensure that this paragraph is not empty

\bibliographystyle{apsrev4-2}
\bibliography{biblio}

\newpage
\newpage{}
\begin{widetext}
\section{Two-level system response}

Close to the excitonic frequency the polarization function Eq.~\eqref{eq:Pi}
has a pole and this is similar to a weakly driven two-level system
at low temperatures. Indeed, consider the Hamiltonian of the two-level
system parametrized by the Pauli matrices $\hat{\sigma}_{x},\hat{\sigma}_{z}$
$H_{\text{TLS}}=d_{\text{TLS}}E(t)\hat{\sigma}_{x}+\frac{\omega_{\text{TLS}}}{2}\hat{\sigma}_{z},$
where $E$ is the external electric field and $d_{\text{TLS}}$ is
the dipole matrix element and $\omega_{\text{TLS}}$ is the transition
energy. Assuming the equilibrium state is $\left|g\right\rangle $,
the response to a weak fluctuating electric field is characterized
by $R_{\text{TLS}}(\omega)=-id_{\text{TLS}}^{2}\int_{0}^{\infty}e^{i\omega t}\left\langle g|\left[\sigma_{x}(t),\sigma_{x}(0)\right]|g\right\rangle $,
computed in Heisenberg picture with respect to $H_{\text{TLS}}$.
Within the single-excitation subspace at zero temperature, we straightforwardly
find $R_{\text{TLS}}(\omega)=-2\omega_{\text{TLS}}d_{\text{TLS}}^{2}/(\omega_{\text{TLS}}^{2}-\left(\omega+i0^{+}\right)^{2})$
thus matching that of our boundary layer close to $\omega_{\text{ex}}$.
We note that while our linear-response treatment is consistent with
the resonant electric susceptibility of TLS, it does not capture the
saturation at large drive and finite temperature. Typically, saturation
requires anharmonicity of the excitonic mode, or equivalently an effective
restriction to the zero- and one-exciton subspace. Such anharmonicity
can arise from the composite nature of the exciton \citep{keldysh2024collective,huang2024nonbosonic},
localization, and exciton-exciton interactions. A microscopic calculation
of this nonlinear response is left for future work.

\section{Analytic solution of Bethe-Salpeter equation}

Here we provide details on how we approach solving Eq.~\eqref{eq:BS}
analytically. For that we use the conventional mapping (row stacking)
the matrix equation onto the vector as follows $\hat{\Gamma}_{3}\rightarrow\vec{\Gamma}_{3}$
with $\vec{\Gamma}_{3}$ (for any pair of operators $O_{\text{r}},O_{\text{l}}$
the mapping is $O_{\text{l}}\hat{\Gamma}_{3}O_{\text{r}}\rightarrow\left(O_{\text{l}}\otimes O_{\text{r}}^{T}\right)\times\vec{\Gamma}_{3}$)
obeying \citep{Roth1934,PetersenPedersen2012}:

\begin{align}
 & \vec{\Gamma}_{3}\left(i\Omega_{m}\right)=\vec{\tau}_{3}+\left(\lambda_{0}\nu_{0}\check{M}(i\Omega_{m})-V^{H}\nu_{0}\vec{\tau}_{3}\vec{\tau}_{3}^{T}\check{M}\right)\times\vec{\Gamma}_{3}(i\Omega_{m}),\label{eq:Gamma3}
\end{align}
where

\[
\check{M}=T\int d\xi_{k}\left(\left(\hat{\tau}_{3}\hat{G}_{{\bf k'}}\left(i\epsilon_{n'}+i\Omega_{m}\right)\right)\otimes\left(\hat{\tau}_{3}\hat{G}_{{\bf k'}}^{T}\left(i\epsilon_{n'}\right)\right)\right)
\]
and $\otimes$ denoting the Kronecker product of matrices, while $\times$
being the ordinary matrix multiplication. This equation can be solved
formally as follows

\begin{equation}
\vec{\Gamma}_{3}\left(i\Omega_{m}\right)=\left(1-\left(\lambda_{0}\nu_{0}\check{M}(i\Omega_{m})-V^{H}\nu_{0}\vec{\tau}_{3}\vec{\tau}_{3}^{\dagger}\check{M}\right)\right)^{-1}\times\vec{\tau}_{3}\label{eq:Gamma_sol}
\end{equation}
We can further use Sherman-Morrison-Woodbury matrix identity to find

\[
\vec{\Gamma}_{3}\left(i\Omega_{m}\right)=\frac{\vec{\Gamma}_{3}^{(0)}}{1+V^{H}\nu_{0}\vec{\tau}_{3}^{\dagger}\times\check{M}\times\vec{\Gamma}_{3}^{(0)}},
\]
where $\vec{\Gamma}_{3}^{(0)}$ denotes the solution without Hartree
shift. In the limit $V^{H}\rightarrow\infty$ we get

\[
\vec{\Gamma}_{3}\approx\frac{\vec{\Gamma}_{3}^{(0)}}{V^{H}\nu_{0}\vec{\tau}_{3}^{\dagger}\times\check{M}\times\vec{\Gamma}_{3}^{(0)}}
\]
and is apparent that the charge sector is completely frozen in this
case, i.e. the bare poles of $\vec{\Gamma}_{3}^{(0)}$generally disappear.

Let us now consider an opposite limit, i.e. when $V^{H}=0$. We now
derive the necessary polarization matrix $\check{M}$. At half-filling
it can be reduced to just 2x2 matrix with $M_{i,j}=\frac{1}{2}\vec{\tau}_{i}^{\dagger}\times\check{M}\times\vec{\tau}_{j}$
with $i,j\in\left[2,3\right]$. This represents the usual charge-phase
coupling in superconductors. Using Eq.~\eqref{eq:G} and taking $T=0$
limit for simplicity, we find:

\begin{align}
M_{2,2} & =\int d\xi_{k}\frac{2\sqrt{\Delta^{2}+\xi_{k}^{2}}}{4\Delta^{2}+4\xi_{k}^{2}+\Omega_{m}^{2}},\label{eq:M22}\\
M_{2,3}=M_{3,2} & =\int d\xi_{k}\frac{\Delta\Omega_{m}}{\sqrt{\Delta^{2}+\xi_{k}^{2}}\left(4\Delta^{2}+4\xi_{k}^{2}+\Omega_{m}^{2}\right)},\label{eq:M23}\\
M_{3,3} & =-\int d\xi_{k}\frac{2\Delta^{2}}{\sqrt{\Delta^{2}+\xi_{k}^{2}}\left(4\Delta^{2}+4\xi_{k}^{2}+\Omega_{m}^{2}\right)},\label{eq:M33}
\end{align}
The integral in Eq.~\eqref{eq:M22} is obviously divergent and it
can be regularized as follows:

\begin{align}
M_{2,2} & =\int d\xi_{k}\left\{ \frac{2\sqrt{\Delta^{2}+\xi_{k}^{2}}}{4\Delta^{2}+4\xi_{k}^{2}+\Omega_{m}^{2}}-\frac{1}{2\sqrt{\Delta^{2}+\xi_{k}^{2}}}\right\} +\int d\xi_{k}\frac{1}{2\sqrt{\Delta^{2}+\xi_{k}^{2}}}\nonumber \\
 & =-\int d\xi_{k}\left\{ \frac{\Omega_{m}^{2}}{2\left(4\Delta^{2}+4\xi_{k}^{2}+\Omega_{m}^{2}\right)\sqrt{\Delta^{2}+\xi_{k}^{2}}}\right\} +\frac{1}{\lambda_{0}\nu_{0}\Delta}.\label{eq:M22_simp}
\end{align}
Where we used the BCS self-consistency equation Eq.~\eqref{eq:delta}
$\Delta=\lambda_{0}\nu_{0}\int d\xi_{k}\frac{\Delta}{2\sqrt{\Delta^{2}+\xi_{k}^{2}}}$.
At this point, it is now convenient to perform the analytic continuation
$\Omega_{m}\rightarrow-i\omega+0^{+}$. Using

\begin{align}
I(\omega) & =\int d\xi_{k}\frac{1}{\sqrt{\Delta^{2}+\xi_{k}^{2}}\left(4\Delta^{2}+4\xi_{k}^{2}-(\omega+i\eta)^{2}\right)}\nonumber \\
 & =\frac{2}{\left(\omega+i\eta\right)\sqrt{4\Delta^{2}-(\omega+i\eta)^{2}}}\arcsin\frac{\omega+i\eta}{2\Delta}\label{eq:I}
\end{align}
At low frequency we get $I(\omega)=\frac{2}{\omega\sqrt{4\Delta^{2}-\omega^{2}}}\arcsin\frac{\omega}{2\Delta}\approx\frac{1}{2\Delta^{2}}$.
Combining Eqs.~(\ref{eq:M22}-\ref{eq:I}) we get:

\[
M_{i,j}=\left(\begin{array}{cc}
\frac{I(\omega)}{2}\omega{}^{2}+\frac{\delta}{\lambda_{0}\nu_{0}\Delta} & -i\omega\Delta I(\omega)\\
-i\omega\Delta I(\omega) & -2\Delta^{2}I(\omega)
\end{array}\right)
\]
As an important benchmark, we expect to have a soft mode for attractive
$\lambda_{0}$ and no proximity. In the limit $\omega\rightarrow0$
and $\delta=\Delta$ we get:

\[
M_{i,j}=\left(\begin{array}{cc}
\frac{1}{\lambda_{0}\nu_{0}} & 0\\
0 & -1
\end{array}\right).
\]
Using Eq.~\eqref{eq:Gamma_sol} we indeed recover a soft mode (i.e.
a zero-frequency pole).

\[
M_{i,j}=\left(\begin{array}{cc}
\frac{I(\omega)}{2}\omega{}^{2}+\frac{\delta}{\lambda_{0}\nu_{0}\Delta} & -i\omega\Delta I(\omega)\\
-i\omega\Delta I(\omega) & -2\Delta^{2}I(\omega)
\end{array}\right).
\]

\subsubsection{Josephson resonance benchmark}

Let us assume that the interaction on metal side is attractive. Expanding
the Bethe-Salpeter pole at low frequency we get:

\[
1-\lambda_{0}\nu_{0}M_{i,j}=\left(\begin{array}{cc}
1-\lambda_{0}\nu_{0}\frac{\omega{}^{2}}{4\Delta^{2}}-\frac{\delta}{\Delta} & \frac{i\omega}{2\Delta^{2}}\Delta\lambda_{0}\nu_{0}\\
\frac{i\omega}{2\Delta^{2}}\Delta\lambda_{0}\nu_{0} & 1+\lambda_{0}\nu_{0}
\end{array}\right)
\]
This yields the Josephson resonance frequency (assuming $\lambda_{0}\nu_{0}\ll1$)

\[
\omega_{J}^{l=0}\approx2\sqrt{\frac{\Delta\Delta_{\text{ind}}}{\lambda_{0}\nu_{0}}}
\]
Obviously, this mode ceases to exist whenever $\lambda_{0}<0$. However,
another mode emerges in this case as we show below. Hartree contribution can be added straightforwardly yielding the Josephson frequency provided in the main text. 

\subsubsection{Repulsion-induced excitonic resonance}

Let us now consider $\lambda_{0}\nu_{0}<0$ case. Determinant of the
matrix (we still assume $\omega<2\Delta$) is:

\begin{align}
 & \det\left(1-\lambda_{0}\nu_{0}\check{M}(i\Omega_{m})\right)=-\frac{\lambda_{0}\nu_{0}\left(4\delta\Delta-4\Delta^{2}+\omega^{2}\right)\arcsin\left(\frac{\omega}{2\Delta}\right)}{\omega\sqrt{4\Delta^{2}-\omega^{2}}}-\frac{\delta}{\Delta}+1\label{eq:Det}
\end{align}
Expanding at $\omega\sim2\Delta$ the subgap poles are controlled
by: 
\[
\frac{\left|\delta\right|\left(\left|\lambda_{0}\right|\nu_{0}+1\right)+\Delta}{\Delta}-\frac{\pi\left|\delta\right|\left|\lambda_{0}\right|\nu_{0}}{2\sqrt{\Delta}\sqrt{2\Delta-\omega}}=0
\]
This yields subgap poles at

\[
\omega_{\text{ex}}\approx2\Delta-\frac{\pi^{2}\delta^{2}\Delta\lambda_{0}^{2}\nu_{0}^{2}}{4(\left|\delta\right|\left(\left|\lambda_{0}\right|\nu_{0}+1\right)+\Delta)^{2}}
\]
At weak coupling we get

\[
\omega_{\text{ex}}\approx2\Delta-\frac{\pi^{2}\delta^{2}\Delta}{4\Delta_{\text{ind}}^{2}}\lambda_{0}^{2}\nu_{0}^{2}
\]

\subsubsection{Finite angular momentum\label{subsec:Finite-angular-momentum}}

For finite angular momentum resonances (See Sec.~\eqref{sec:Angular-decomposition}),
in the absence of any Fermi surface distortions, the $\hat{M}$ matrix
remains the same. Accordingly, determinant is found straightforwardly:
\[
\det\left(1-\lambda_{l}\nu_{0}\check{M}(i\Omega_{m})\right)=\left\{ 4\Delta^{2}-\omega^{2}-4\delta\Delta\lambda_{l}\nu_{0}\right\} \frac{\lambda_{l}\nu_{0}\arcsin\left(\frac{\omega}{2\Delta}\right)}{\omega\sqrt{4\Delta^{2}-\omega^{2}}}-\frac{\delta\lambda_{l}}{\Delta\lambda_{0}}+1
\]
And the corresponding bound state energy at weak coupling is (a more
complete expression is given in Eq.~\eqref{eq:lneq} in the main
text)

\[
\omega_{\text{ex}}^{l}\approx2\Delta-\frac{\pi^{2}\delta^{2}\Delta}{4\Delta_{\text{ind}}^{2}}\frac{\lambda_{l}^{4}}{\lambda_{0}^{2}}\nu_{0}^{2}.
\]

\subsubsection{Hartree contribution}

We now estimate the shift of the resonance in the presence of Hartree
term. Due to its structure Eq.~\eqref{eq:Gamma_sol} it just modifies
the matrix as

\begin{align*}
 & \lambda_{0}\nu_{0}\check{M}(i\Omega_{m})-V^{H}\nu_{0}\vec{\tau}_{3}\vec{\tau}_{3}^{\dagger}\check{M}\rightarrow\\
 & \left(\begin{array}{cc}
\lambda_{0}\nu_{0}\frac{I(\omega)}{2}\omega{}^{2}+\frac{\delta}{\Delta} & -\lambda_{0}\nu_{0}i\omega\Delta I(\omega)\\
-\left(\lambda_{0}\nu_{0}-2V^{H}\nu_{0}\right)i\omega\Delta I(\omega) & -2\left(\lambda_{0}\nu_{0}-2V^{H}\nu_{0}\right)\Delta^{2}I(\omega)
\end{array}\right)
\end{align*}
The resonance equation becomes at weak coupling becomes

\[
\frac{\delta}{\Delta}(\left(\lambda_{0}-2V^{H}\right)\nu_{0}-1)+2V^{H}\nu_{0}+1\approx\frac{\pi\nu_{0}(\delta\lambda_{0}+2V^{H}\Delta_{\text{ind}})}{2\sqrt{\Delta}\sqrt{2\Delta-\omega_{\text{res}}}}
\]

\begin{align*}
\omega_{\text{ex}} & =2\Delta-\frac{\pi^{2}\Delta(\delta\lambda_{0}\nu_{0}+2V^{H}\nu_{0}\Delta_{\text{ind}})^{2}}{4(\delta\lambda_{0}\nu_{0}+(1+2V^{H}\nu_{0})\Delta_{\text{ind}})^{2}}
\end{align*}
We note that the existence of the pole is only guaranteed once $|\delta||\lambda_{0}|-2|V^{H}|\Delta_{\text{ind}}>0$.

\subsubsection{Pole residue}

It is also interesting to explore the pole residue at the resonance.
Let us start without the Hartree contribution. We will be interested
in expanding $(1-\lambda_{0}\nu_{0}\check{M}(\omega+i\eta))^{-1}$
at $\omega\sim\omega_{\text{res}}$. Using Eq.~\eqref{eq:Det} we
get

\begin{align*}
 & \Lambda=\det\left(1-\lambda_{0}\nu_{0}\check{M}(\omega+i\eta)\right)^{-1}\\
 & \approx\frac{\pi^{2}\delta^{2}\Delta^{2}\lambda\nu^{2}}{2(\delta(\lambda\nu-1)+\Delta)^{3}\left(\omega+i0^{+}-\omega_{\text{ex}}\right)}\\
 & \approx\frac{\pi^{2}\delta^{2}\Delta^{2}\lambda\nu^{2}}{2\Delta_{\text{ind}}^{3}\left(\omega+i0^{+}-\omega_{\text{ex}}\right)}
\end{align*}
Close to the band edge we find:

\begin{align*}
 & \Pi(\omega)\sim\nu_{0}\vec{\tau}_{3}^{\dagger}\times\check{M}\times\vec{\Gamma}_{3}^{(0)}\\
 & \approx\nu_{0}\frac{\pi^{2}\delta^{2}\Delta^{2}\lambda_{0}^{2}\nu_{0}^{2}}{2\Delta_{\text{ind}}^{3}\left(\omega+i0^{+}-\omega_{\text{ex}}\right)}\left\{ \frac{\Delta_{\text{ind}}}{\Delta}+\frac{\pi\Delta_{\text{ind}}}{2\sqrt{\Delta}\sqrt{2\Delta-\omega_{\text{ex}}}}+\ldots\right\} 
\end{align*}
At weak coupling this simplifies to

\[
\Pi(\omega)\approx\nu_{0}\frac{\pi^{2}\delta\lambda_{0}\nu_{0}}{2\left(\omega+i0^{+}-\omega_{\text{ex}}\right)}
\]
The $s$-wave exciton residue is thus given by $Z_{\text{ex}}=\pi^{2}\delta\lambda_{0}\nu_{0}/2$.

\[
\]

\section{Angular decomposition\label{sec:Angular-decomposition}}

In this section we consider the case when the rotational symmetry
of the system is broken and the finite-center-of-mass response. Furthermore,
we demonstrate that the result in the most general case can be formally
represented in a very compact form, similar to Eq.~\eqref{eq:Gamma_sol}.
Let us start with Bethe-Salpeter equation Eq.~\eqref{eq:BS} provided
in the main text and perform the angular decomposition. Here we also
consider a more general scenario: finite center-of-mass momentum ${\bf q}$.
We note however that the same approach is easily extendable to other
scenarios such as involving rotational symmetry breaking (see Subsec.~\ref{subsec:Rotational-symmetry-breaking}).
The density response at finite ${\bf q}$ is described by

\begin{align}
 & \hat{\Gamma}_{3}\left({\bf k},{\bf q},i\Omega_{m}\right)=\hat{\tau}_{3}+\frac{T}{{\cal A}}\sum_{{\bf k}',n'}V_{{\bf k-k'}}\hat{\tau}_{3}\hat{G}_{{\bf k'+\frac{q}{2}}}\left(i\epsilon_{n'}+i\Omega_{m}\right)\hat{\Gamma}_{3}({\bf k}',{\bf q},i\Omega_{m})\hat{G}_{{\bf k'-\frac{q}{2}}}(i\epsilon_{n'})\hat{\tau}_{3}\nonumber \\
 & -\frac{T}{{\cal A}}V_{{\bf q}}^{H}\tau_{3}\sum_{{\bf k}',n'}\text{Tr}\hat{\tau}_{3}\hat{G}_{{\bf k'+\frac{q}{2}}}\left(i\epsilon_{n'}+i\Omega_{m}\right)\hat{\Gamma}_{3}({\bf k}',{\bf q},i\Omega_{m})\hat{G}_{{\bf k'-\frac{q}{2}}}(i\epsilon_{n'}),\label{eq:Gamma3q}
\end{align}
where we now allow the Green's function $\hat{G}_{{\bf k}}$ to depend
on both the magnitude of $k$ and its orientation. Projecting both
sides onto $l'$ angular channel we get

\begin{align}
 & \hat{\Gamma}_{3}\left(k,l',{\bf q},i\Omega_{m}\right)\equiv\int\frac{d\phi_{k}}{2\pi}e^{-i\phi_{k}l'}\hat{\Gamma}_{3}\left({\bf k},{\bf q},i\Omega_{m}\right)=\nonumber \\
 & \hat{\tau}_{3}\delta_{l',0}+T\int\frac{kdk'}{2\pi}\frac{d\phi_{k'}}{2\pi}\int\frac{d\phi_{k}}{2\pi}e^{-i\phi_{k}l'}V_{{\bf k-k'}}\hat{\tau}_{3}\hat{G}_{{\bf k'+\frac{q}{2}}}\left(i\epsilon_{n'}+i\Omega_{m}\right)\sum_{l''}e^{i\phi_{k'}l''}\hat{\Gamma}_{3}(k',l'',{\bf q},i\Omega_{m})\hat{G}_{{\bf k'-\frac{q}{2}}}(i\epsilon_{n'})\hat{\tau}_{3}\nonumber \\
 & -\delta_{l',0}TV_{{\bf q}}^{H}\tau_{3}\sum_{n'}\int\frac{k'dk'}{2\pi}\frac{d\phi_{k'}}{2\pi}\text{Tr}\hat{\tau}_{3}\hat{G}_{{\bf k'+\frac{q}{2}}}\left(i\epsilon_{n'}+i\Omega_{m}\right)\sum_{l''}e^{i\phi_{k'}l''}\hat{\Gamma}_{3}(k',l'',{\bf q},i\Omega_{m})\hat{G}_{{\bf k'-\frac{q}{2}}}(i\epsilon_{n'})\label{eq:Gamma3-1}
\end{align}
With our decomposition of the interaction term $V_{{\bf k-k'}}=\sum_{m}\lambda_{m}e^{im\left(\phi_{{\bf k}}-\phi_{{\bf k'}}\right)}$
the renormalized vertex $\hat{\Gamma}_{3}$ is independent of the
magnitude of $k$ (which will be dropped as an argument in the following).
Our goal is now to represent Eq.~\eqref{eq:Gamma3-1} as a simplified
matrix equation. Using row stacking we get:

\begin{align*}
\\
\vec{\Gamma}_{3}\left(l',{\bf q},i\Omega_{m}\right)= & \vec{\tau}_{3}\delta_{l',0}+T\lambda_{l'}\sum_{l''}\check{M}_{l',l''}\times\check{\Gamma}_{3}(l'',{\bf q},i\Omega_{m})-\delta_{l',0}TV_{{\bf q}}^{H}\sum_{l''}\vec{\tau}_{3}\vec{\tau}_{3}^{T}\times\check{M}_{0,l''}\times\hat{\Gamma}_{3}(l'',{\bf q},i\Omega_{m})
\end{align*}
where 
\begin{align*}
\check{M}_{l',l''} & =T\sum_{n'}\int\frac{k'dk'}{2\pi}\int\frac{d\phi_{k'}}{2\pi}e^{-i(l'-l'')\phi_{k'}}\left(\left(\hat{\tau}_{3}\hat{G}_{{\bf k'+\frac{q}{2}}}(i\epsilon_{n'}+i\Omega_{m})\right)\otimes\left(\hat{\tau}_{3}\hat{G}_{{\bf k}'-\frac{{\bf q}}{2}}^{T}(i\epsilon_{n'})\right)\right)
\end{align*}
It is now apparent that we can also represent the angular component
in the vectorized form

\begin{align*}
 & \vec{\Gamma}_{3}\left({\bf q},i\Omega_{m}\right)\equiv\vec{\tau}_{3}\left|0\right\rangle +(\lambda\star\check{M})\times\check{\Gamma}_{3}({\bf q},i\Omega_{m})-V_{{\bf q}}^{H}\left(\left|0\right\rangle \vec{\tau}_{3}\vec{\tau}_{3}^{T}\left\langle 0\right|\right)\times\check{M}\times\hat{\Gamma}_{3}({\bf q},i\Omega_{m}),
\end{align*}
where $\left|l\right\rangle $ represents the vector of angular components
and $(\lambda\star\check{M})_{l,l'}\equiv\lambda_{l}\check{M}_{l,l'}$.
Again, we can formally solve this equation and get:

\begin{equation}
\vec{\Gamma}_{3}\left({\bf q},i\Omega_{m}\right)\equiv\left((1-(\lambda\star\check{M})+V_{{\bf q}}^{H}\left(\left|0\right\rangle \vec{\tau}_{3}\vec{\tau}_{3}^{T}\left\langle 0\right|\right)\times\check{M}\right)^{-1}\times\vec{\tau}_{3}\left|0\right\rangle .\label{eq:Gamma3Full}
\end{equation}
Using this expression, the full polarization function can be written
as

\[
\Pi({\bf q},i\Omega_{m})=\left\langle 0\right|\vec{\tau}_{3}\times\check{M}\times\vec{\Gamma}_{3}\left({\bf q},i\Omega_{m}\right)
\]

\subsubsection{Higher $l$ solutions}

Assuming rotationally symmetric Fermi surface, we find that there
are no contributions to \eqref{eq:Gamma3Full} at ${\bf q}=0$. In
this case the $\check{M}$ matrix is not mixing angular momentum harmonics.
However, other excitons are still valid solutions. Indeed assuming
we perturb the system with some $L\neq0$ angular momentum drive,
we would get:

\[
\vec{\Gamma}'_{3}\left({\bf q},i\Omega_{m}\right)\equiv\left((1-(\lambda\star\check{M})\right)^{-1}\times\vec{\tau}_{3}\left|L\right\rangle 
\]
This equation is solved in SubSec.~\eqref{subsec:Finite-angular-momentum}.

\subsection{Analytic continuation}

For numerical purposes, it is convenient to analytically continue
the $M$ matrix. Using spectral decomposition of BDG Green's functions
we get:

\begin{align*}
\check{M}_{l',l''} & =T\sum_{n'}\int\frac{k'dk'}{2\pi}\int\frac{d\phi_{k'}}{2\pi}e^{-i(l'-l'')\phi_{k'}}\hat{\tau}_{3}\hat{G}_{{\bf k'+\frac{q}{2}}}\left(i\epsilon_{n'}+i\Omega_{m}\right)\otimes\hat{\tau}_{3}\hat{G}_{{\bf k'-\frac{q}{2}}}^{T}(i\epsilon_{n'})\\
 & =T\sum_{n'}\int\frac{k'dk'}{2\pi}\int\frac{d\phi_{k'}}{2\pi}e^{-i(l'-l'')\phi_{k'}}\hat{\tau}_{3}\int\frac{dx}{\pi}\frac{\hat{g}_{{\bf k'+\frac{q}{2}}}\left(x\right)}{x-i\epsilon_{n'}-i\Omega_{m}}\otimes\hat{\tau}_{3}\int\frac{dy}{\pi}\frac{\hat{g}_{{\bf k'-\frac{q}{2}}}^{T}(y)}{y-i\epsilon_{n'}},
\end{align*}
where $\hat{g}_{{\bf k}}(x)=\text{Im}\hat{G}_{{\bf k}}^{R}(x)$. Using
that $\hat{g}_{{\bf k}}(x)=-\pi\sum_{s}\delta\left(x-sE_{{\bf k}}\right)\hat{P}_{s}(\xi_{{\bf k}})$,
where $E_{{\bf k}}$ is the BDG band dispersion and $\hat{P}_{s}(\xi_{{\bf k}})=\frac{1}{2}\left\{ \hat{\tau}_{0}+s\frac{\xi_{{\bf k}}\hat{\tau}_{3}+\hat{\tau}_{1}\Delta}{\sqrt{\Delta^{2}+\xi_{{\bf k}}^{2}}}\right\} $
is the projector onto this band. Evaluating the Matsubara sum, we
get:

\begin{align*}
 & M_{l',l''}^{a,b}(i\Omega_{m})=\frac{\vec{\tau}_{a}^{T}\times\check{M}_{l',l''}\times\vec{\tau}_{b}}{2}\\
 & =\frac{1}{2}\int\frac{k'dk'}{2\pi}\int\frac{d\phi_{k'}}{2\pi}e^{-i(l'-l'')\phi_{k'}}\sum_{s,s'}\text{Tr}\left\{ \hat{\tau}_{a}\hat{\tau}_{3}\hat{P}_{s}(\xi_{{\bf k'+\frac{q}{2}}})\hat{\tau}_{b}\hat{P}_{s'}(\xi_{{\bf k'-\frac{q}{2}}})\hat{\tau}_{3}\right\} \frac{\tanh\left(\frac{\beta s'E_{{\bf k'-\frac{q}{2}}}}{2}\right)-\tanh\left(\frac{\beta sE_{{\bf k'+\frac{q}{2}}}}{2}\right)}{2sE_{{\bf k'+\frac{q}{2}}}-2s'E_{{\bf k'-\frac{q}{2}}}-2i\text{\ensuremath{\Omega}m}}
\end{align*}
At zero temperature and replacing $i\Omega_{m}\rightarrow\omega+i0^{+}$
we get:

\begin{equation}
M_{l',l''}^{a,b}(\omega+i0^{+})=\sum_{s}\int\frac{k'dk'}{2\pi}\int\frac{d\phi_{k'}}{2\pi}e^{-i(l'-l'')\phi_{k'}}\frac{\text{Tr}\left\{ \hat{\tau}_{a}\hat{\tau}_{3}P_{s}(\xi_{{\bf k'+\frac{q}{2}}})\hat{\tau}_{b}P_{-s}(\xi_{{\bf k'-\frac{q}{2}}})\hat{\tau}_{3}\right\} }{2}\frac{\text{sign}\left(s\right)}{(\omega+i0^{+})-s\left(E(\xi_{{\bf k'+\frac{q}{2}}})+E(\xi_{{\bf k'-\frac{q}{2}}})\right)}\label{eq:Mabll}
\end{equation}

\subsection{Finite momentum q}

Eq.~\eqref{eq:Mabll} and Eq.~\eqref{eq:Gamma3Full} can be evaluated
numerically by linearizing the dispersion, i.e. $\xi_{{\bf k\pm\frac{q}{2}}}\approx\xi_{{\bf k}}\pm\frac{v_{F}q}{2}\cos\phi_{kq}$,
where $\phi_{kq}$ is the angle between ${\bf k}$ and ${\bf q}$
vectors. The required integrals can then be taken straightforwardly.
The results are shown in Fig.~\ref{Fig1}~(b) of the main text.

\subsection{Rotational symmetry breaking\label{subsec:Rotational-symmetry-breaking}}

We model the rotational symmetry breaking by assuming the electron
masses are different along $x$ and $y$ directions. More specifically,
we assume the electronic dispersion is

\begin{align*}
\xi_{{\bf k}} & =\frac{k_{x}^{2}}{2m_{x}}+\frac{k_{y}^{2}}{2m_{y}}-\mu=\frac{k^{2}}{2}\left\{ \frac{\cos^{2}\phi_{k}}{\left(m+\delta m\right)}+\frac{\sin^{2}\phi_{k}}{\left(m-\delta m\right)}\right\} -\mu\\
 & \approx\frac{k^{2}}{2m}\left\{ 1-\frac{\delta m}{m}\cos2\phi_{k}\right\} -\mu
\end{align*}
where we assumed $\delta m/m\ll1$. We note that here we consider
the simplest case of mass modification while more complex scenarios
involving spin components will be presented elsewhere. Numerically,
the density response can now be found straightforwardly using Eqs.~(\ref{eq:Gamma3q},
\ref{eq:Gamma3-1}, \ref{eq:Gamma3Full}).
\end{widetext}

\end{document}